# Probing the surface of silicon and the silicon-silica interface using nonperturbative third and fifth harmonic generation

J. Seres[1,*], E. Seres[1], E. Céspedes[2,3] L. Martinez-de-Olcoz[2], M. Zabala[2], T. Schumm[1]

[1]Atominstitut - E141, Technische Universität Wien, Stadionallee 2, 1020 Vienna, Austria

[2]Instituto de Microelectrónica de Barcelona, IMB-CNM (CSIC), Campus UAB, Cerdanyola, 08193, Barcelona, Spain

[3]Present address: Instituto de Ciencia de Materiales de Madrid (CSIC), Cantoblanco. 28049 Madrid, Spain

We examined, in backward (reflection) geometry, the generation of the 3rd and 5th harmonics, located in the deep and vacuum ultraviolet, on the surface of silicon and on the interface between silicon and silica when a thin silica film was grown on a silicon substrate. In both cases, a strong dependence of the harmonic signal on the polarization direction of the driving laser beam was found. The differences observed for both samples, are qualitatively explained. Furthermore, a simplified tensor formalism for the polarization dependence is introduced, which reveals the structural symmetry of the surface and the interface and describes the polarization dependence with high accuracy. The study is an essential step to further understand nonlinear interaction and nonperturbative harmonic generation on the boundaries of materials.

**Introduction**

Converting the wavelength of the ultrashort pulses of laser frequency combs to the ultraviolet (UV) and vacuum ultraviolet (VUV) wavelengths makes them potential sources for several new applications. It can be used for high-resolution electronic transition spectroscopy, to measure new high energy atomic transitions with high precision and to develop next-generation atomic clocks. The ultrashort pulses in the UV and VUV wavelengths regime would be useful for tracking fast chemical and biological processes. High harmonic generation (HHG) has been proven to be a suitable technique for the conversion to shorter wavelengths, reaching wavelength even below 100 nm. In recent years, different solid crystalline materials demonstrated the potential to replace noble gases as nonlinear medium for frequency conversion, avoiding the technical difficulties accompanied with high gas loads and differential pumping of vacuum chambers.

Among the practical solid media, silicon (Si) was tested successfully in recent years. Silicon is a key material in the semiconductor industry and widely used in modern information technology. It is available in very high quality and has high damage threshold [1], which is an important factor for HHG: to reach good conversion efficiency, high laser intensities in the GW/cm$^2$ or even in the TW/cm$^2$ are required.

Beyond being a suitable medium for the generation of the optical harmonics, silicon is also a prime material to investigate and further understand the microscopic aspects of the process of harmonic generation. In many applications of Si, electrophysical processes on the surface and at the interface between Si and other materials play an important role. Non-destructive inspection of Si was demonstrated using 3[rd] harmonic generation [2]. In materials with inversion symmetry such as Si, odd order harmonics can be generated effectively. With the development of new laser sources, inspection of internal and inter-layer defects became possible using 3rd and 5th harmonic generation [3]. These results suggest that generating different orders of harmonics can be a suitable tool for non-contact, non-destructive, high-resolution, and depth-selective inspection of Si wafers.

Obviously, generation of harmonics in Si even at high orders is in the interest of several research groups recently and it is studied experimentally and theoretically [4, 5]. Experiments typically used laser pulses in the infrared, at wavelength of 1.3 µm to 3.8 µm, below the bandgap in which Si is transparent and the duration of the used laser pulses was in the range of 50 fs up to 400 fs [3, 6-8]. Harmonics were observed in transmission (forward) geometry and also appeared in reflection from the samples (backward geometry). The harmonic spectrum extending from the visible even into the extreme ultraviolet spectral range was successfully generated in silicon crystals. As a promising direction, forming nanostructures, nano-antennas on the surface of Si, plasmon-assisted harmonics were produced to enhance the conversion efficiency or to lower the required laser intensities [9-11].

In this study, 3rd and 5th harmonics are generated on the surface of Si wafer samples in reflection geometry using 20 fs pulses of a Ti:sapphire frequency comb (108 MHz) at 800 nm, where the Si is not transparent. The obtained source is a deep ultraviolet (DUV) and VUV frequency comb that can be applied for high precision measurements in these new wavelengths. Furthermore, harmonic generation on a thin film of $SiO_2$ grown onto the Si substrate is investigated.

**Description of harmonic generation on a boundary**

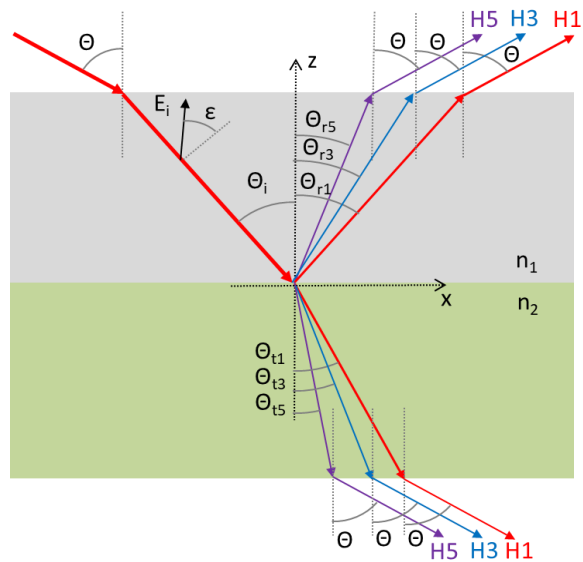

Fig. 1: Generation of harmonics on the boundary of two materials with refractive indexes of $n_1$ and $n_2$. Red arrows represent the fundamental laser beam while the blue and violet arrows are the 3rd and 5th harmonic beams in forward (t: transmitted) and backward (r: reflected) directions.

To understand the measurement results and to design a suitable experimental setup, here a short description of the harmonic generation at the boundary of two nonlinear materials is given. The examined case can be described with an example of plan-parallel slabs with a boundary between them as presented in Fig. 1. We are interested in the nonlinear interaction of the laser light with the boundary and the generation of the harmonics at this boundary. Such configuration can be realized when a substrate is coated with a thin film and the boundary is the interface between the substrate and the film. A specific case is when the first medium (thin film) is vacuum, and the boundary is the surface of a material itself.

In this article, the interaction on the boundary is handled similarly to Ref. [12] and the Snell's law of refraction is applied for all derivations. The laser beam arrives at the boundary with angle of incidence (AOI) $\Theta_i$ defined by

$$n_{11} sin\theta_i = sin\theta. \tag{1}$$

Here, $n_{11} = n_1(\omega_1)$ or in general: $n_{ij} = n_i(\omega_j)$. The incident laser pulses polarize the boundary, which becomes the source of new beams. The laser intensity is high enough to make the polarization nonlinear, therefore on the nonlinear boundary, beyond the linear response yielding the reflected and refracted laser beams, additional nonlinear contributions, namely the harmonics of the laser beam, emerge. The order of a harmonic is denoted by $q$ with an angular frequency $\omega_q = q\omega_1$ and obviously for the fundamental laser beam $q$ = 1. For the reflected laser beam $\theta_{r1} = \theta_i$ and for the refracted (transmitted) beam $n_{11} sin\theta_i = n_{21} sin\theta_{t1}$. The harmonic beams generated at the boundary can propagate both in forward and in backward direction, with directions that are different from the fundamental beams to fulfill the phase matching condition for the wave vectors

$$(\vec{k}_q - q\vec{k}_1)\vec{r} = 0. \tag{2}$$

The coordinate system is chosen, such that the boundary is the surface at z = 0, the plane of incidence is at y = 0, and lies in the xz plane. Under these conditions $\vec{k}_q = (k_{qx}, 0, k_{qz})$, $\vec{r} = (x, y, 0)$ and consequently, only the tangential components define the phase matching, namely $k_{qx} = qk_{1x}$. Applying it to the harmonic beams on both sides of the boundary provides us with the equations:

$$n_{1q} sin\theta_{rq} = n_{11} sin\theta_{r1} = n_{11} sin\theta_i \tag{3}$$

$$n_{2q} sin\theta_{tq} = n_{21} sin\theta_{t1} = n_{11} sin\theta_i. \tag{4}$$

From Eq. (3) and (4), using Eq. (1), one obtains

$$n_{1q} sin\theta_{rq} = n_{2q} sin\theta_{tq} = sin\theta. \tag{5}$$

Eq. (5) defines the directions of the generated harmonic beams in the backward and forward directions. They propagate in different directions inside the mediums while the refractive indices of the materials are different for the different harmonic orders. However, after leaving the mediums by refracting on the surfaces, both the backward and forward harmonic beams co-propagate with the fundamental laser beam, as presented in Fig. 1. Such co-propagation was mentioned in Ref. [13] and studied in detail in Ref. [14] as a proof of the surface/interface origin of the generated harmonics.

**Experimental setup**

To generate harmonics on the surface of Si, the measurement arrangement of Fig. 2 was used. The setup is designed by considering that the backward harmonic beams co-propagate with the reflected laser beam, as described in the previous section. The laser source was a Ti:sapphire frequency comb (FC8004, Menlo Systems). It delivered laser pulses at repetition rate of 108 MHz and with pulse duration of 18 fs at the central wavelength of 800 nm. The carrier offset frequency and the repetition rate can be stabilized and the later scanned when performing high precision spectroscopic measurements. The typical pulse energy was 8 nJ, which was focused onto the surface of the sample using a lens with focal length of 10 mm. It produced a beam waist of 6.0±0.5 µm, and an on-axis peak intensity of 1.0±0.2 TW/cm². In some measurements, a lens with focal length of 15 mm was also used resulting in a beam waist of 9.0±0.5 µm and an on-axis peak intensity of 0.45±0.1 TW/cm².

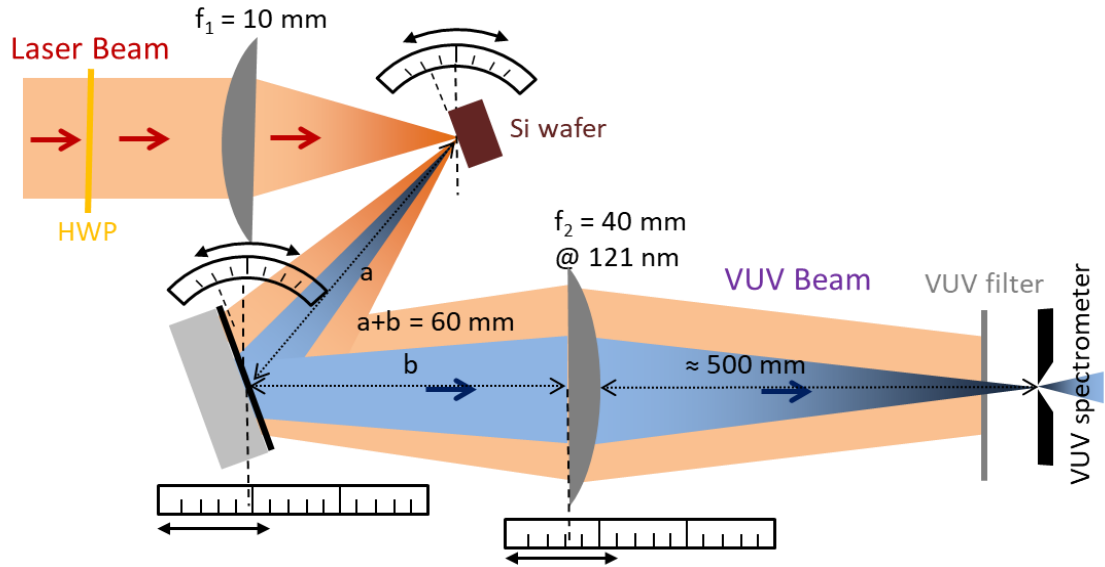

Fig. 2: Experimental setup. The laser beam was focused on the surface of a Si wafer and the generated 5$^{th}$ harmonic was focused on the input slit of the VUV spectrometer. The direction of the polarization was set by a half-wave-plate (HWP).

The reflected laser beam together with the generated beams of harmonics were directed to the entrance of the VUV spectrometer (McPherson 234/302) using a VUV mirror. Both, the sample and the mirror, were placed onto rotation stages to hold them parallel while measurements were performed at different angles of incidence. The distance (≈ 60 mm) of the VUV-grade MgF$_2$ lens from the sample was set to focus the generated 5$^{th}$ harmonic beam to the input slit of a VUV spectrometer which was located at about 500 mm from the lens. Under this condition, the 3$^{rd}$ harmonic and especially the fundamental laser beam is not focused to the input slit, giving a better contrast for the measurements by decreasing the scattered background in the spectrometer. Furthermore, it was possible to insert a VUV bandpass filter (160-BB-1D, Pelham Research Optical L.L.C.) to further suppress the fundamental and 3$^{rd}$ harmonic beam. The spectrally resolved beam was detected with a VUV photomultiplier (Hamamatsu R6836), sensitive in the 115-320 nm spectral range. The setup of the source and the VUV spectrometer were in vacuum with a background pressure of 10$^{-3}$ mbar to avoid the reabsorption of the 5$^{th}$ harmonic by air. Both the VUV mirror and the lens were placed onto translation stages to ensure the focusing of the 5$^{th}$ harmonic beam into the input slit of the spectrometer at different AOIs.

**Generation of harmonics on the surface of silicon**

Using the experimental setup presented in Fig. 2, harmonics were generated in two sample configurations. In the first one, a Si substrate was used ($n_2$ = 3.694 @ 800 nm [15]). The Si (100) substrate characteristics were P-type (boron), 10-20 Ωcm, 500-µm-thick, 2 sides-polished. Here the generation of harmonics was studied directly on the vacuum-silicon boundary ($n_1$ = 1). In the second case, a SiO$_2$ film was grown on the Si substrate, and the harmonics were generated at the SiO$_2$ – Si boundary ($n_1$ = 1.461 @ 800 nm [16]). The SiO$_2$ layer was thermally grown at 1100°C (experimental thickness was 389 ± 1 nm, measured by optical reflectometry (Nanospec 6100)). This thermally grown SiO$_2$ layer is typically an amorphous layer (more specifically considered as a vitreous non-crystalline solid) with a stoichiometric relation between Si and O, except in the first few nanometers (less than 10 nm) of interface with the underlying silicon substrate. Prior to $SiO_2$ thermal growth, both Silicon substrates underwent a chemical cleaning procedure (HF plus piranha solution) to remove the native silicon oxide. Since no further chemical etching was performed preceding optical measurements (some days after), differing from the 400 nm thermally grown stoichiometric $SiO_2$, a very thin amorphous silicon oxide is expected onto the Si (100) substrate. Previous studies have shown that silicon is one of few materials whose native oxide will self-limit its growth at a thickness of about 2 nm in a matter of hours [17]. Furthermore, this native amorphous oxide onto Si (100) exhibits a varying composition with depth, gradually changing from a highly oxidized, near stoichiometric state at the surface to a silicon rich phase near the interface. High depth resolution medium energy ion scattering (MEIS) spectrometry analysis showed an outer oxide layer of about 0.7 nm of stoichiometric $SiO_2$, and an underlying sub-oxide $SiO_x$ one (x < 2) of approximately 0.6 nm [18].

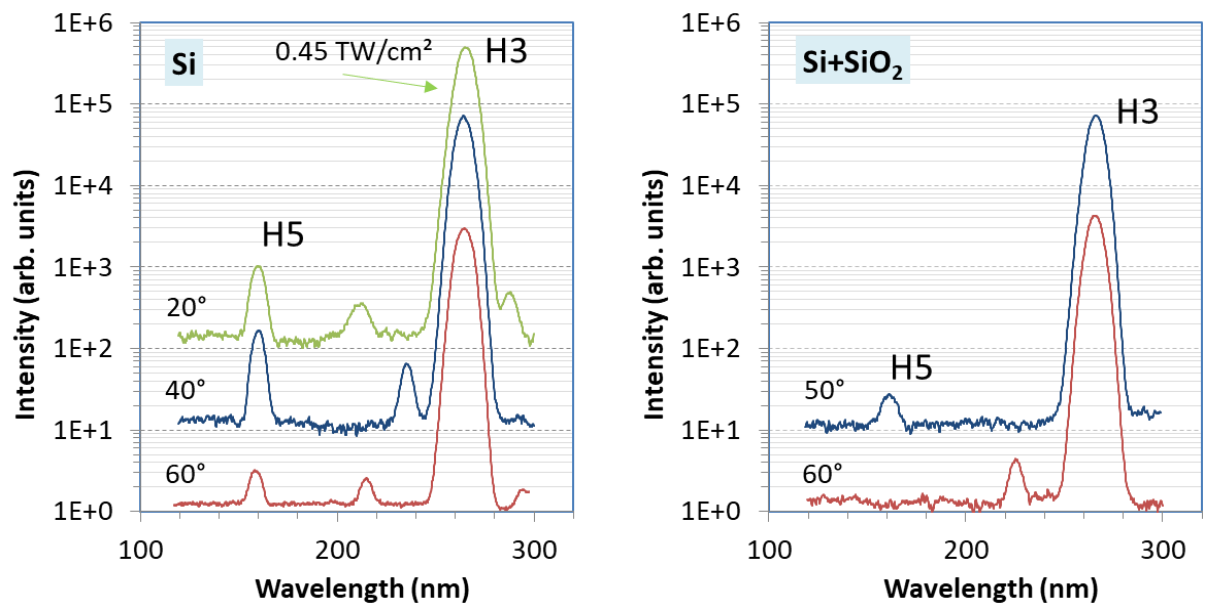

Fig. 3. Measured spectra of the harmonics generated on the pure surface of the Si substrate and when a $SiO_2$ thin film was grown on the substrate. Reflected (backward) harmonics were measured at different angles of incidence and at peak laser intensity of 1 $TW/cm^2$ except one at 20° on Si.

The spectra of generated harmonics have been recorded at different angles of incidence and are presented in Fig. 3. The possible values of the AOIs were limited: Using a focusing lens with f = 10 mm, it was not possible to set angles below 40° because the reflected beams hit the lens. The upper value was limited to 60° because the distance between the sample and the VUV lens should have been about 60 mm to image the sample onto the input slit of the VUV spectrometer. Therefore, for the measurement at 20°, a lens with f = 15 mm was used. For all spectral measurements, the polarization direction of the linearly polarized laser beam was set to be parallel to the plane of incidence ($\varepsilon = 0$), when the intensity of the generated harmonics was maximal. From the surface of the pure Si, in Fig. 3, one can observe the generation of strong third harmonic (H3, 267 nm, 4.6 eV) and the weaker fifth harmonic (H5, 160 nm, 7.75 eV). The intensity of both decreases with the increase of the AOI ($\theta$), consistent with other observations [7, 14]. A third peak between the harmonics was also observed, sometimes only at shorter wavelength than H3 and sometimes at both sides of H3. The origin of these peaks is not fully clarified; it can be an anti-Stokes/Stokes electronic Raman scattering of H3 between a higher state in the conduction band of the exited Si and a lower one. Their distance from H3 changes up to 1.1 eV. In a material with inversion symmetry, such as silicon, only odd harmonics can occur. It is consistent with our observation, because even harmonics as H2, H4 and H6, which would be in the observable range of the VUV spectrometer, have not appeared.

Further, we examined the effect of a thin $SiO_2$ film grown onto the Si substrate on the generation of harmonics. Crystalline $SiO_2$ has high band gap of 8.1 eV [19], so it would be fully transparent even at the wavelength of H5 (7.75 eV). In our case, the $SiO_2$ film was amorphous as mentioned above. Amorphous $SiO_2$ has a smaller bandgap of 7.5 eV [19] meaning an absorption α = 1.4 µm$^{-1}$ [16], and consequently the film absorbs about half of the H5 signal.

From the presence of the film, one can expect contradictory behaviors. On one hand, the actual AOI at the Si-$SiO_2$ boundary is smaller, $\theta_i$ ≈ 32° and 36° in the case of θ ≈ 50° and 60°, which predicts larger harmonic signal according to the previous measurement. On the other hand, earlier measurements [14, 20] obtained smaller harmonic signal from the interface of two materials compared to the material – vacuum surfaces. To address this question, we measured spectra for a Si substrate containing about 400-nm-thick $SiO_2$ film, shown in Fig. 3. The obtained spectra look somewhat different from the pure Si case. The 3$^{rd}$ harmonics are somewhat stronger (less than 2-times); however, the 5$^{th}$ harmonics are weaker, and even at θ = 60°, it is below the detection threshold. It was not possible to perform a measurement at θ = 40° because the sample was overheated and damaged within the time interval of the spectral measurement even when using a lens with a longer f = 15 mm focal length.

**Polarization dependence of the harmonic signal**

To gain information about the silicon-vacuum and silicon-silica interface, the polarization direction (ε) of the linearly polarized incident laser beam was rotated using a half-wave plate (Fig. 2), and at few AOIs, the intensity of the generated H3 and H5 was measured at 267 nm and 160 nm, respectively. The measurement results are presented in Fig. 4.

Considering that Silicon is a crystalline material with cubic $O_h$ symmetry (point group m3m), the third-order susceptibility tensor has 21 nonzero elements [21]. To simplify the calculation of the polarization dependence of the H3 signal, we introduced a 2-dimensional susceptibility tensor (3x3) instead of the original 4-dimensional one, a simplification which is possible for this crystal symmetry (and also for several others like isotropic, cubic, orthorhombic, few hexagonal and few tetragonal crystals):

$$\begin{aligned}
\chi_{xx}^{(3)} &= \chi_{xxxx}^{(3)} & \chi_{xy}^{(3)} &= \chi_{xxyy}^{(3)} + \chi_{xyxy}^{(3)} + \chi_{xyyx}^{(3)} & \chi_{xz}^{(3)} &= \chi_{xxzz}^{(3)} + \chi_{xzxz}^{(3)} + \chi_{xzzx}^{(3)} \\
\chi_{yx}^{(3)} &= \chi_{yyxx}^{(3)} + \chi_{yxyx}^{(3)} + \chi_{yxxy}^{(3)} & \chi_{yy}^{(3)} &= \chi_{yyyy}^{(3)} & \chi_{yz}^{(3)} &= \chi_{yyzz}^{(3)} + \chi_{yzyz}^{(3)} + \chi_{yzzy}^{(3)} \\
\chi_{zx}^{(3)} &= \chi_{zzxx}^{(3)} + \chi_{zxzx}^{(3)} + \chi_{zxxz}^{(3)} & \chi_{zy}^{(3)} &= \chi_{zzyy}^{(3)} + \chi_{zyzy}^{(3)} + \chi_{zyyz}^{(3)} & \chi_{zz}^{(3)} &= \chi_{zzzz}^{(3)}
\end{aligned} \quad (6)$$

The linearly polarized electric field with polarization angle ε and incidence angle $\theta_i$ can be written in the general form:

$$E(\omega) = E_0(\omega) \begin{pmatrix} e_x \\ e_y \\ e_z \end{pmatrix} = E_0(\omega) \begin{pmatrix} \cos(\varepsilon)\cos(\theta_i) \\ \sin(\varepsilon) \\ \cos(\varepsilon)\sin(\theta_i) \end{pmatrix}. \quad (7)$$

The generated H3 field than can be expressed in a simple form for m3m point group:

$$\begin{pmatrix} E_x(3\omega) \\ E_y(3\omega) \\ E_z(3\omega) \end{pmatrix} \propto \chi_{11}^{(3)} E_0^3(\omega) \begin{pmatrix} e_x \\ e_y \\ e_z \end{pmatrix} \circ \begin{pmatrix} 1 & a & a \\ a & 1 & a \\ a & a & 1 \end{pmatrix} \begin{pmatrix} e_x^2 \\ e_y^2 \\ e_z^2 \end{pmatrix}. \quad (8)$$

Silicon – Vacuum boundary

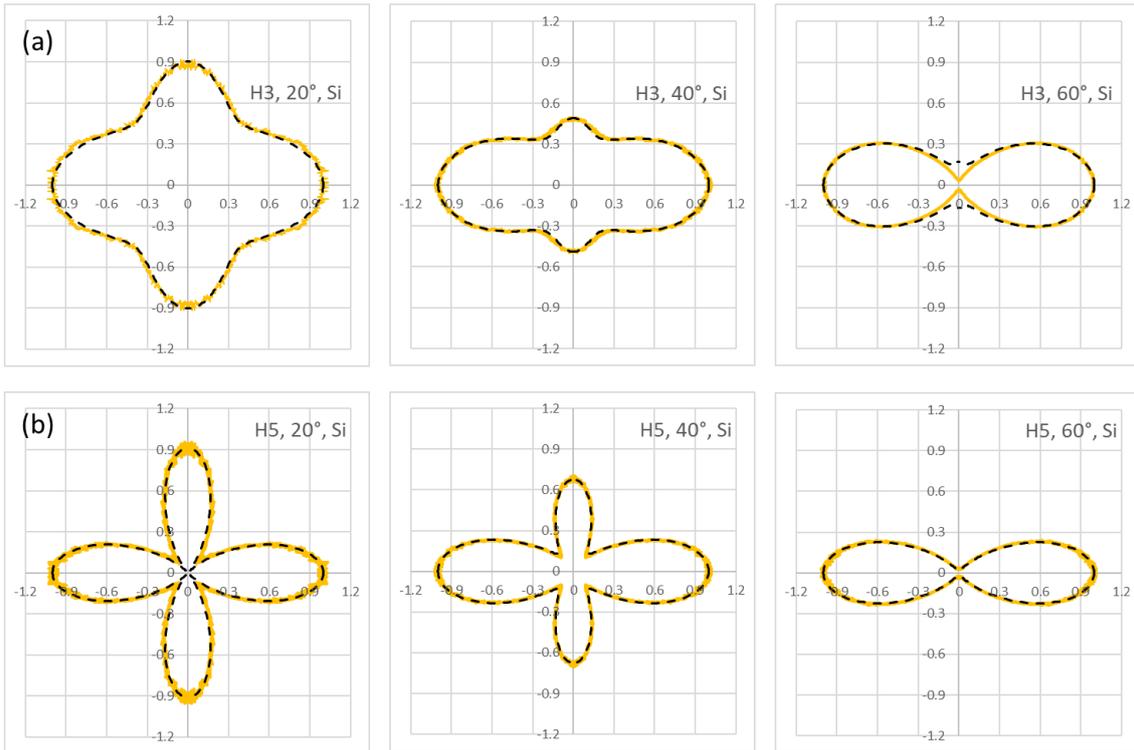

Silicon – Silica boundary

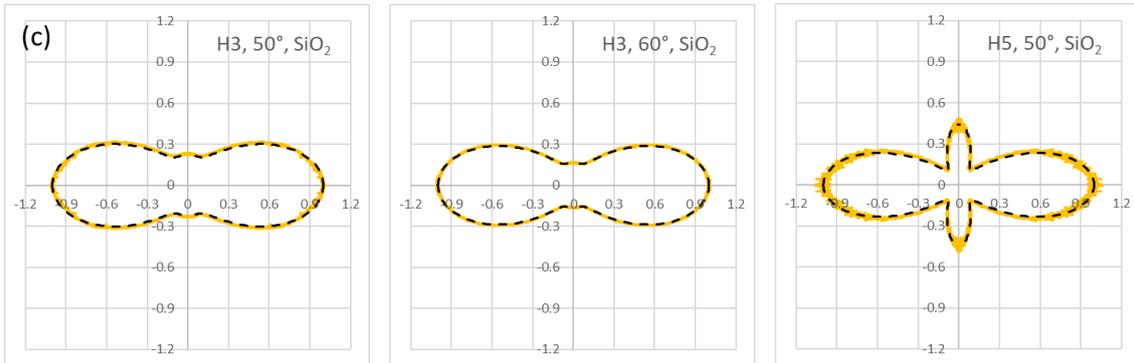

Fig. 4: The 3$^{rd}$ and 5$^{th}$ harmonic signals are strongly dependent on the direction of the polarization of the driving laser beam for both cases when the harmonics are generated (a, b) on the silicon surface or (c) on the silicon-silica interface. Yellow: measurement; black-dashed: theory. The details, how theory curved were calculated can be found in the text.

In Eq. (8), the first column vector on the right side means element-wise multiplication. According to the literature [22], for silicon $a$ = 1.27. However, one must consider that the laser beam is a transversal wave, and the generated harmonic beam should be also transversal. Taking the example of the silicon surface, the harmonic beam leaves the surface at the same angle as the incidence angle ($\Theta_r = \Theta_i$) and for the beams of the laser and of the backward harmonics: $|E_z/E_x| = \tan(\theta_i)$. It means that the first and the last rows of the tensor in Eq. (8) should be the same (or multiplied by -1), meaning $a$ = 1. It means an isotropic medium and in isotropic media (such as gases), H3 and other odd harmonics can be routinely generated. Independently, the polarization dependence of H3 was calculated for Si at several values of "$a$" and compared to the measurement in Fig. 4(a) and no similarity has been found. It can be concluded that the surface has a different symmetry compared to the Si (bulk) crystal. The surface can be an orthorhombic $C_{2v}$ (mm2) symmetry (justification will be given in the next chapter) and the same type of formalism as Eq. (8) can be applied but with independent tensor elements. A similar formalism is also suitable to calculate H5. In general, for harmonic $q = 3$ and $q = 5$:

$$\begin{pmatrix} E_x(q\omega) \\ E_y(q\omega) \\ E_z(q\omega) \end{pmatrix} \propto E_0^r(\omega) \begin{pmatrix} e_x \\ e_y \\ e_z \end{pmatrix} \circ \begin{pmatrix} \chi_{xx}^{(q)} & \chi_{xy}^{(q)} & \chi_{xz}^{(q)} \\ \chi_{yx}^{(q)} & \chi_{yy}^{(q)} & \chi_{yz}^{(q)} \\ \chi_{zx}^{(q)} & \chi_{zy}^{(q)} & \chi_{zz}^{(q)} \end{pmatrix} \begin{pmatrix} e_x^{r-1} \\ e_y^{r-1} \\ e_z^{r-1} \end{pmatrix}. \quad (9)$$

In Eq. (9), for generalization, "$r$" is used as the rank of the nonperturbative process. As it has been concluded in several publications [23-27], "$r$" should not be necessarily equal to the harmonic order "$q$" and it should not be integer eighter. Eq. (9) was successfully used to describe the polarization dependence of the harmonics, and the calculated curves (black dashed) in Fig. 4 were obtained by using the next tensors:

Silicon-Vacuum (SV), Fig. 4(a, b), for H3 $r = 4$ and for H5 $r = 5$:

$$\chi_{SV}^{(3)} = \chi_{11}^{(3)} \begin{pmatrix} 1 & 1 & 3.2 \\ 1.4 & 0.91 & 4.5 \\ 1 & 1 & 3.2 \end{pmatrix}, \qquad \chi_{SV}^{(5)} = \chi_{11}^{(5)} \begin{pmatrix} 1 & 0 & -7 \\ -1 & 0.7 & 9 \\ 1 & 0 & -7 \end{pmatrix}, \quad (10)$$

Silicon-Silica (SS), Fig. 4(c), for H3 $r = 3$ and for H5 $r = 5$:

$$\chi_{SS}^{(3)} = \chi_{11}^{(3)} \begin{pmatrix} 1 & 1 & 5 \\ 1 & 1 & 0 \\ 1 & 1 & 5 \end{pmatrix}, \qquad \chi_{SS}^{(5)} = \chi_{11}^{(5)} \begin{pmatrix} 1 & 0 & 13 \\ 4 & 1 & 0 \\ 1 & 0 & 13 \end{pmatrix}. \quad (11)$$

Using these values, good agreement between the calculated and measured curved in Fig. 4 has been found.

**Discussion**

Measuring and calculating the polarization-dependent harmonic signals for H3 and H5 provides the opportunity to understand the underlying processes, structure, and symmetry of the silicon surface and the silicon-silica interface. Looking at the tensors of Eq. (10) and Eq. (11), few features can be observed:

(i) The tensors in Eq. (10) and Eq. (11) immediately show that the $C_{2v}$ symmetry of the pure silicon surface ($\chi_{11}^{(q)} \neq \chi_{22}^{(q)} \neq \chi_{33}^{(q)}$) changed to a $C_{4v}$ symmetry when accounting for the silica film ($\chi_{11}^{(q)} = \chi_{22}^{(q)} \neq \chi_{33}^{(q)}$). For explanation, Fig. 5 visualizes the crystal structures using the software of Ref. [28]. Fig. 5(a) shows the silicon crystal surface (001) from above. Atoms represented in dark blue color are on the surface and others in lighter blue are below the surface. Taking all the atoms, the silicon crystal has a tetragonal 4-fold symmetry from this direction. Similarly, the atoms only on the surface suggest also a 4-fold $C_{4v}$ symmetry. The crystal surface is illuminated by the laser beam along the atomic chains in the x-direction. Looking, however, at the surface from the side in Fig. 5(b), one can recognize that the atoms on the surface are chemically bonded to the underlaying atoms only in the x-directions (highlighted in yellow dashed lines) and bonds in the y-direction appear only between the lower atomic layers inside the crystals. Atoms inside the crystal are bonded to 4 other Si atoms with 3($sp^3$) tetrahedral hybrid bonds, albeit the atoms on the surface bonded only to 2 other Si atoms and probably have $3s^23p^2$ electronic structure from which the two 3p orbitals are bonded. It means that the atoms and their bonds on the surface has only a 2-fold $C_{2v}$ symmetry. This can explain the obtained Eq. (10) tensors showing a $\chi_{xx}^{(3)}/\chi_{yy}^{(3)} = 1.1$ and $\chi_{xx}^{(5)}/\chi_{yy}^{(5)} = 1.4$ while the atoms on the surface can be easier polarized in the directions of the bonds (x-direction) than in the direction perpendicular to them (y-direction).

The formation of such a surface can happen as a consequence of the laser illumination. The surface is illuminated from x-direction under certain AOIs. The intensity of the laser is high enough to remove the native, few atomic layers thick oxidized layer easily an even can remove the upper Si atomic layer if the bonds are not directed in the direction of the illumination and a surface as represented in Fig. 5(b) is produced. The surface gets a $C_{2v}$ symmetry. Furter removal of atomic layers and degradation of the surface was not observed even after illumination for longer time.

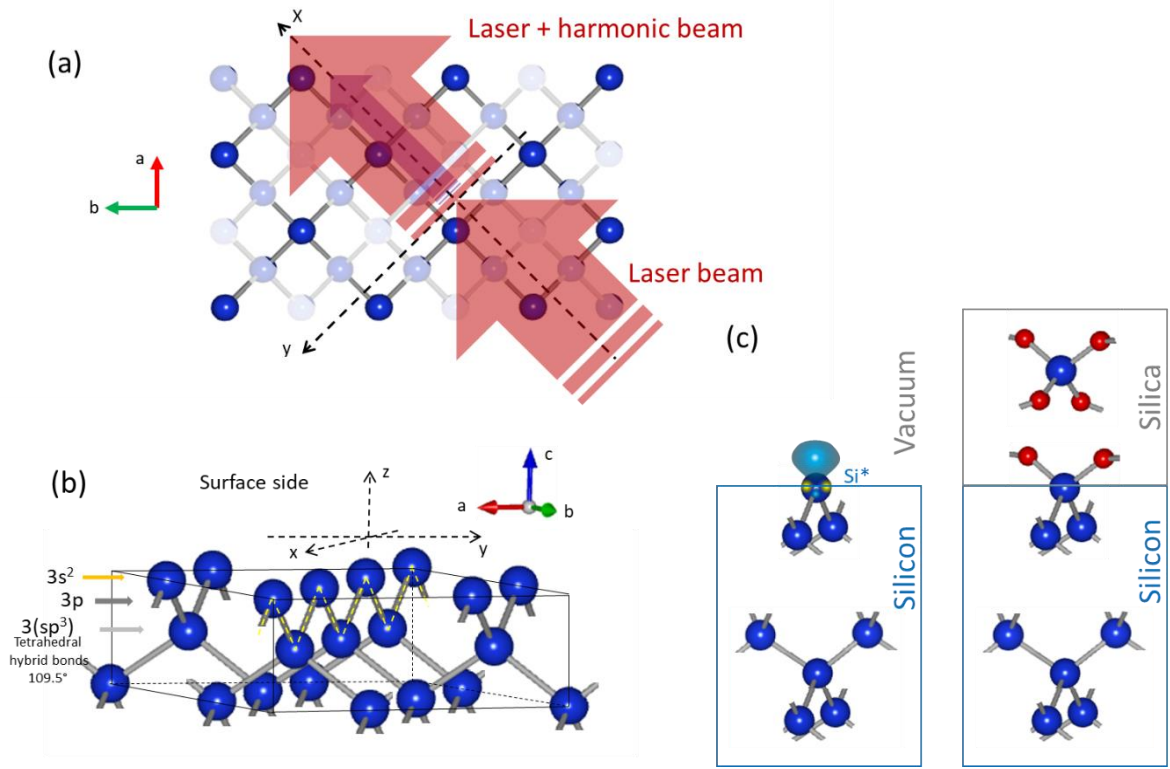

Fig. 5: (a) The silicon surface is illuminated by the laser beam within the xz plane of incidence and harmonics are generated. The 4-fold symmetry of the Si atoms (dark blue on the surface, light blue under the surface) are well visible. (b) Side-view of the crystal surface. (c) Visualization of the difference between the silicon-vacuum and silicon-silica boundary.

(ii) Another interesting observation is that, for H3 and especially for H5, the polarizability of the atoms in the z-direction is the dominant one. To understand this, the building blocks (not unit cells) of the silicon crystal are separately represented in Fig. 5(c). As it has been mentioned, inside the crystal, silicon atoms are bonded to 4 neighbors, and on the surface, the silicon atoms are bonded to 2 neighbors from the lower layer. The remaining electrons of the atoms on the surface can be excited by the laser (photon energy of 1.54 eV is higher than the 1.2 eV bandgap of silicon) and form an out-of-surface orbital, which is represented with light blue color in Fig. 5(c). The formation of such orbitals was calculated [29] on the surface of $Al_2O_3$ and described with conduction-band-edge wave functions. Such orbitals were used [14] to understand the generation of harmonics on the surface of AlN. The out-of-surface orbital is oriented perpendicular to the silicon surface and consequently can be easily polarized in z-direction. This might explain the large polarizability and large tensor values in the z-direction.

(iii) The interface has a C$_{4v}$ symmetry, when a 400-nm-thick SiO$_2$ film was thermally grown on the Si substrate. In Eq. (11), the tensor became more symmetric compared to Eq. (10), namely the different main axes components $\chi_{11}^{(3)} \neq \chi_{22}^{(3)}$ and $\chi_{11}^{(5)} \neq \chi_{22}^{(5)}$ along the pure silicon surface changed to symmetric, $\chi_{11}^{(3)} = \chi_{22}^{(3)}$ and $\chi_{11}^{(5)} = \chi_{22}^{(5)}$ for silica-silicon interface. The symmetry of the boundary changed from C$_{2v}$ to a more symmetric C$_{4v}$.

To understand the difference, the two cases are visualized in Fig. 5(c). Inside the silicon, the Si atoms are bonded to 4 neighboring Si atoms with tetrahedral hybrid bonds. Inside the silica, the Si atoms are bonded to 4 neighboring oxygen atoms also with tetrahedral hybrid bonds. The oxygen atoms are bonded to 2 neighboring Si atoms. The essential difference can be observed at the boundary. When the silica film is present on the silicon surface, out-of-surface orbitals of unbonded electrons are not able to form and are replaced with tetrahedral hybrid bonds with two oxygen atoms. Under the same conditions (same laser pulses and focusing), we generated harmonics on the surface of a silica substrate and observed only weak H3 and no H5. It means that the Si-O bonds can be hardly polarized, which can be expected from the high 7.5 eV bandgap of the silica. Therefore, the easily polarizable out-of-surface orbitals disappear and are replaced with hardly polarizable Si-O bonds, which obviously determines the susceptibility tensors and tetrahedral bond structure on the boundary means more symmetric polarizability in x- and y-directions. Furthermore, it can predict/explain the lower harmonic intensities on this interface compared to the harmonics generated on the surface of silicon.

(iv) Another difference can be observed in the second row of the tensors, which describes the polarization parallel with the boundary, in y-direction. While, in the case of the silicon surface, the laser field component perpendicular to the surface strongly affected the harmonic generation ($\chi_{23}^{(3)}$ and $\chi_{23}^{(5)}$ were the largest tensor elements), in the case of silicon-silica interface, these tensor elements became zero. It is not straightforward to analyze, why the out-of-surface orbitals cause strong coupling between the y- and z-directions. It would require a quantum-mechanical calculation of these orbitals and their interaction with the laser filed.

(v) The ranks of the nonperturbative H3 generation are different for the two scenarios. This observation was checked by directly measuring the dependence of the harmonic intensities on the laser intensity. For the Si surface, $I(3\omega) \propto a \cdot I^4(\omega) + b \cdot I^{1.95}(\omega)$ and $I(5\omega) \propto I^5(\omega)$, furthermore for the Si-SiO$_2$ interface $I(3\omega) \propto a \cdot I^3(\omega) + b \cdot I^{1.9}(\omega)$ has been found, which for high intensities gives the same rank as obtained from the polarization measurements in Fig. 4. Because of the smaller signal, for H5 on the Si-SiO$_2$ interface, it was not possible to perform a reliable measurement. An explanation can be that in the case of the Si-SiO$_2$ interface, a static electric field is built up at the boundary [8], which can contribute to the generation of the harmonics. We must note here that the measurements of polarization dependence, like in Fig. 4, support a more accurate determination of the rank of the nonperturbative nonlinear process than the direct measurement of the intensity dependence.

**Summary**

We examined, in backward (reflection) geometry, the generation of the 3$^{rd}$ and 5$^{th}$ harmonics on the surface of silicon and on the interface between silicon and silica when a thin silica film was grown on a silicon substrate. In both cases, a strong dependence of the harmonic signals on the polarization direction of the driving laser beam was observed. Furthermore, a strong dependence of the shape of the polarization curves on the angle of incidence of the laser beam was found. The differences, observed in the harmonic signals, were qualitatively explained by the formation of the out-of-surface electron orbitals on the surface of the silicon, and the formation of tetrahedral bonds at the boundary between the silicon and silica. Furthermore, a simplified tensor formalism of the polarization dependence was introduced, which determined the symmetry of the surface and interface and described the polarization dependence with accuracy.

In the same geometry, stronger 3$^{rd}$ harmonics and weaker 5$^{th}$ harmonics were generated on the silicon-silica interface than on the silicon surface. The weaker observed 5$^{th}$ harmonics might be the consequence of the reabsorption of the signal in the silica film. To explore this contradiction, further studies are necessary with films of different thickness, material, and crystallization quality.

**Acknowledgement**

This work is part of the ThoriumNuclearClock project that has received funding from the European Research Council (ERC) under the European Union's Horizon 2020 research and innovation programme (Grant agreement No. 856415). Work has been performed within the project 20FUN01 TSCAC, which has received funding from the EMPIR programme co-financed by the Participating States and from the European Union's Horizon 2020 research and innovation programme. We acknowledge support from the Österreichische Nationalstiftung für Forschung, Technologie und Entwicklung (AQUnet project). This research has used the Spanish ICTS Network "MICRONANOFABS", partially funded by FEDER funds through "MINATEC-PLUS-2" project FICTS2019-02-40.